\pdfoutput=1 
\documentclass[12pt]{article}

\usepackage{inputenc}

\usepackage{amsmath,amssymb,bbm,color}
\usepackage{amsbsy}
\usepackage{fixmath}
\usepackage{euscript}
\usepackage{graphicx}
\usepackage{hyperref}
\usepackage{cite}
\usepackage{enumerate}
\numberwithin{equation}{section}

\newcommand{\msbar}{\overline{\text{MS}}}

\newcommand{\mc}{\text{MC}}
\newcommand{\lo}{\text{LO}}
\newcommand{\nlo}{\text{NLO}}

\newcommand{\GeV}{\text{GeV}}

\newcommand{\veps}{{\varepsilon}}

\newcommand{\Dmf}{\mathfrak{D}}

\newcommand{\krknlo}{{\textsf{KrkNLO}}}

\newcommand{\herwig}[1]{\textsf{Herwig #1}}

\newcommand{\mcatnlo}[1]{\textsf{MC@NLO#1}}

\newcommand{\powheg}[1]{\textsf{POWHEG#1}}

\begin{document}

\begin{titlepage}

\begin{flushright}
\bf IFJPAN-IV-2016-9\\
    CERN-TH-2016-132 \\
    MCnet-16-18
\end{flushright}

\vspace{5mm}
\begin{center}
    {\Large\bf Parton distribution functions \vspace{2mm}\\
               in Monte Carlo factorisation scheme$^{\star}$ }
\end{center}

\vskip 10mm
\begin{center}
{\large S.\ Jadach$^a$, W.\ P\l{}aczek$^b$, S.\ Sapeta$^{c,a}$, 
        A.\ Si\'odmok$^{c,a}$ and M.\ Skrzypek$^a$ }

\vskip 2mm
{\em $^a$Institute of Nuclear Physics, Polish Academy of Sciences,\\
  ul.\ Radzikowskiego 152, 31-342 Krak\'ow, Poland}\\
\vspace{1mm}
{\em $^b$Marian Smoluchowski Institute of Physics, Jagiellonian University,\\
ul.\ {\L}ojasiewicza 11, 30-348 Krak\'ow, Poland}\\
\vspace{1mm}
{\em $^c$Theoretical Physics Department, CERN, Geneva, Switzerland }

\end{center}

\vspace{5mm}
\begin{abstract}
\noindent
A next step in development of the \krknlo{} method 
of including complete NLO QCD corrections 
to hard processes in a LO parton-shower Monte Carlo (PSMC) is presented.
It consists of generalisation of
the method, previously used for the Drell--Yan process,
to Higgs-boson production.
This extension is accompanied with the complete description of
parton distribution functions (PDFs) in a dedicated, Monte Carlo (MC) factorisation scheme,
applicable to any process of production of one or more colour-neutral particles
in hadron--hadron collisions.
\end{abstract}

\vspace{2mm}

\vspace{15mm}

\vspace{50mm}
\footnoterule
\noindent
{\footnotesize
$^{\star}$This work is partly supported by 
 the Polish National Science Center grant DEC-2011/03/B/ST2/02632 and
 the Polish National Science Centre grant UMO-2012/04/M/ST2/00240.
}

\end{titlepage}

\newpage

\newpage
\section{Introduction}

The method 
of including complete NLO QCD corrections 
to hard processes in the LO parton-shower Monte Carlo (PSMC), nicknamed \krknlo{}, 
was originally proposed in  Ref.~\cite{Jadach:2011cr},
where its first numerical implementation on top of a toy-model
PSMC was also presented.
It was restricted there to gluon emission only and was elaborated
for two processes: $Z/\gamma^*$ production in hadron--hadron collisions, i.e. the Drell--Yan (DY) process
and deep inelastic electron--hadron scattering (DIS).

In the following Ref.\cite{Jadach:2015mza}, the \krknlo{} method was implemented
for  $Z/\gamma^*$ production process at large hadron collider (LHC)
in combination with Sherpa \cite{Gleisberg:2008ta} 
and Herwig++~\cite{Bahr:2008pv,Bellm:2013lba,Gieseke:2012ft} PSMCs.
Many NLO-class numerical results 
(distributions of transverse momenta, rapidity, integrated cross sections, etc.)
were presented there and comparisons 
of the \krknlo{} predictions with those from other methods, such as \mcatnlo{} \cite{Frixione:2002ik}
and \powheg{} \cite{Nason:2004rx}, were also performed.

The main advantage of the \krknlo{} method with respect to other, older
methods of matching the fixed-order NLO calculations with PSMCs (\mcatnlo{} and \powheg{} ) is its simplicity.
This simplicity stems from the fact that the entire NLO corrections
are implemented using a simple positive multiplicative MC weight.
However, in order to profit from it, 
one has to use in the \krknlo{} method parton distribution functions (PDFs) in a special,
so-called Monte Carlo (MC) factorisation scheme and PSMC has to fulfil some minimum quality criteria.
Most of modern PSMCs are good enough for the \krknlo{} method.

Construction of PDFs in the MC factorisation scheme (FS) has evolved step by step:
in Ref.~\cite{Jadach:2011cr} it was defined for gluonstrahlung only
(albeit for two different processes, DY and DIS).
In Ref.\cite{Jadach:2015mza}, the \krknlo{}  PDFs in the MC FS were defined
and numerically constructed including also gluon to quark transitions/splittings,
relevant for the complete NLO corrections in the DY process, which at the LO
level has only quarks and antiquarks in the initial state.
PDFs in the MC scheme in Ref.~\cite{Jadach:2015mza} were defined in terms
of the standard $\msbar$ PDFs, and constructed numerically by transforming
the $\msbar$ PDFs into MC-scheme PDFs, before they were plugged into PSMC
used in the \krknlo{} method.

However, in Ref.~\cite{Jadach:2015mza} certain elements 
in the transition matrix $K$, transforming the $\msbar$ PDFs into the MC-scheme PDFs 
could be omitted,
because they were not relevant (i.e.\ of a NNLO class) for the DY process.
These elements of the transition matrix have to be added for any process with
initial-state gluons, such as the Higgs-boson production elaborated in the present work.
They will be defined and applied in the following, 
such that the complete transition matrix $K$ transforming 
the $\msbar$ PDFs into the MC-scheme PDFs will be specified
for the first time.
It will be argued that PDFs in such a MC-scheme can serve in the \krknlo{}
method for any process at a hadron--hadron collider in which a colour-neutral
single or multiple system of heavy particles is produced.
For other processes, with one or more coloured partons in the final state
at LO level, the \krknlo{} method with PDFs in the MC scheme may also work,
but this subject is reserved for the forthcoming publications.

The MC factorisation scheme is a complete scheme, such that NLO coefficient functions
for any hard process under consideration are known, 
hence PDFs in the MC FS can be fitted directly to experimental DIS and DY data.
However, at present, we obtain them from PDFs in the $\msbar$ scheme
and leave out direct fitting to data for the future developments.

On the methodological side, as seen in Refs.~\cite{Jadach:2011cr,Jadach:2015mza},
the essence of the \krknlo{} method is that certain NLO correction terms
in an unintegrated/exclusive form present in the $\msbar$ scheme,
which are proportional to unphysical Dirac-delta
terms in transverse momentum of emitted real partons, 
are removed in the \krknlo{} methodology by means of redefinition of PDFs
from  the $\msbar$ to MC scheme.
These `pathological' terms are preventing the use of a simple multiplicative
MC weight for implementing NLO corrections in the $\msbar$ scheme
in real-emission phase space, and they complicate implementation
of the \mcatnlo{} and \powheg{} methods.
These peculiar terms can be determined and calculated either by means of
studying the
NLO corrections to hard process (coefficient functions), or, alternatively,
by means of integrating soft-collinear counter terms 
(similar to these in the Catani--Seymour method \cite{Catani:1996vz}),
which define the MC-scheme PDFs in $d=4+2\veps$ dimensions%
\footnote{They also form matrix elements of the $K$-matrix transforming 
 PDFs from the $\msbar$ to MC scheme.}.
We are going to calculate them using both methods, obtaining the same results.

Last but not least, the NLO calculations for the DY process of Ref.\cite{Jadach:2015mza}
were also compared with NNLO calculations, concluding that they are closer
to the latter than the results of the \mcatnlo{} and \powheg{} methods.

The outline of the paper is the following:
in Section 2 the \krknlo{} method is characterised briefly.
In Section 3 all distributions needed for  implementation of 
the \krknlo{} method for Higgs-boson production in gluon--gluon fusion are elaborated, 
including also many analytical crosschecks 
and a necessary update of the virtual corrections
in soft-collinear counter terms  
used in Ref.\cite{Jadach:2015mza} for the $Z/\gamma^*$ (DY) process.
Section 4 presents numerical results for PDFs in the MC scheme.
Then, first numerical results for the total
cross section from the \krknlo{} method
for the Higgs production at the LHC are shown in Section~5.
Finally, in Section 6 we summarise the paper and discuss future prospects of our work.

\section{The method}
\label{sec:method}

The \krknlo{} method was formulated in a few variants.
For instance, in the version of Ref.~\cite{Jadach:2011cr}, the MC weight implementing
the NLO corrections sums the contributions from all relevant partons
generated in PSMC next to the hard process ``democratically'',
such that it works equally well for PSMCs based on angular ordering 
or virtuality ordering,
contrary to \powheg{} which requires adding extra gluons to a PSMC event.
In the present work, we are going to follow the variant of \krknlo{} 
discussed in Ref.\cite{Jadach:2015mza},
in which the NLO-correcting MC weight uses only one parton, 
the one closest to the hard process in the transverse momentum, 
that is the 1$^\text{st}$ parton generated in the backward evolution (BEV) 
in the PSMC algorithm with $k_T$-ordering.

In any case, in the \krknlo{} method, the entire event of PSMC is preserved
and reweighted, contrary to \powheg{} and \mcatnlo{} where the parton attributed to the hard
process is generated outside PSMC 
and, only later on, the remaining partons are provided by PSMC.
Obviously, this puts certain minimum quality requirements on the the PSMC:
(i) the 1$^\text{st}$ parton in BEV algorithm has to be generated with the distribution which
has a correct soft and collinear limit and
(ii) its phase space in momentum and flavour space
has to be covered completely, without empty regions.
Luckily, the above requirement is fulfilled by all modern PSMCs
for initial-state emissions discussed in this work.

It is worth to comment in advance on the apparent
use in the following of the soft-collinear counter terms (dipoles)
of the Catani--Seymour (CS) subtraction scheme \cite{Catani:1996vz}.
Their role is two-fold: 
(1) the CS dipoles serve us as a useful benchmark, 
as they provide a reference model for QCD distributions of real emissions
featuring the exact soft and collinear limits and
(2) the CS scheme helps us in a proper inclusion of the NLO virtual corrections.
However, let us point out immediately an important difference between the MC and CS scheme:
the CS dipoles do not include virtual corrections,
while soft-collinear counter terms (SCCTs) of the \krknlo{} do include them,
albeit not calculated from Feynman diagrams, 
but deduced from PDF momentum sum rules.
The role of the SCCTs in the \krknlo{} methodology is also much richer than that of the dipoles
in the CS scheme --
our SCCTs not only provide subtractions of soft-collinear singularities 
in real-emission phase space, but they are also used to define PDFs 
in the MC factorisation scheme.
Moreover, their sums are required to coincide with 
the corresponding sums of real-parton distributions in PSMC%
\footnote{At least for the initial-state emitters in the present work,
but also in the final-state ones in the future implementations of the \krknlo{} method.
In fact, SCCTs of the \krknlo{} and PSMC distributions do not need to coincide exactly,
but optional additional weight bringing the PSMC 
to SCCT distribution of the \krknlo{} method has to be well behaved.
}.

\section{Higgs production in gluon--gluon fusion}
\label{sec:MCscheme}

In the following we are going to collect all distributions needed
for implementation of the \krknlo{} method for 
the gluon-fusion Higgs production in hadron--hadron collisions.
Elements of the matrix transforming PDFs from the $\msbar$ to MC scheme
will be also obtained as a byproduct.

\begin{figure}[h]
  \centering
  \includegraphics[width=90mm]{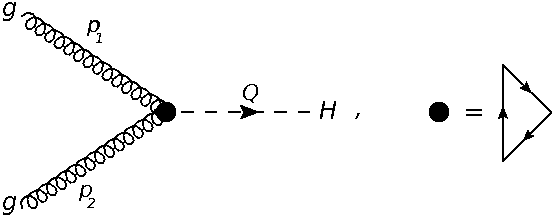}
  \caption{\sf
    The LO Feynman diagram for the process of Higgs boson production 
    in gluon--gluon fusion. 
    The effective vertex (black dot) corresponds to a quark loop with summation
    over all quarks, in which the top-quark mass is set to infinity while the 
    masses of the other quarks are set to zero.
    }
  \label{fig:ggH}
\end{figure}

We start necessarily from the leading order (LO) process
\begin{equation}
g(p_1)\; +\; g(p_2) \:\longrightarrow\: H(Q)\,, 
\label{eq:LOproc}
\end{equation}
see Fig.~\ref{fig:ggH}, where $Q = p_1 + p_2$. 
The LO matrix element squared, in the limit $m_t \rightarrow \infty$ 
and neglecting all other quarks contributions, reads
\begin{equation}
|{\cal M}_{gg}^{\lo}|^2 = \frac{\alpha_s^2}{576\pi^2 v^2} \,Q^4,
\label{eq:LOME}
\end{equation}
where $v^2 = (\sqrt{2}G_F)^{-1}$ is the Higgs vacuum expectation value (VEV) squared.
Hence, the LO total cross section takes the form
\begin{equation}
\sigma_0 \equiv \sigma_{gg}^{\lo}(Q^2) =  \frac{\pi}{Q^4} |{\cal M}_{gg}^{\lo}|^2 =
\frac{\alpha_s^2}{576\pi v^2}\,. 
\label{eq:sigtot0}
\end{equation}

For all NLO subprocesses (channels)
\begin{equation}
a(p_1)\; +\; b(p_2) \:\longrightarrow\: H(Q) + c(k), 
\label{eq:NLOproc}
\end{equation}
where $a$ and $b$ are incoming partons (gluons and/or quarks), 
while $c$ is an outgoing parton (quark or gluon)
we shall use the same parametrisation of the kinematics
in terms of the following Sudakov variables:
\begin{equation}
\alpha = \frac{p_2\cdot k}{p_1\cdot p_2},\qquad 
\beta  = \frac{p_1\cdot k}{p_1\cdot p_2},
\qquad \alpha + \beta = 1 - z \leq 1.
\label{eq:SudVar}
\end{equation}

\begin{figure}[!ht]
  \centering
  \includegraphics[width=100mm]{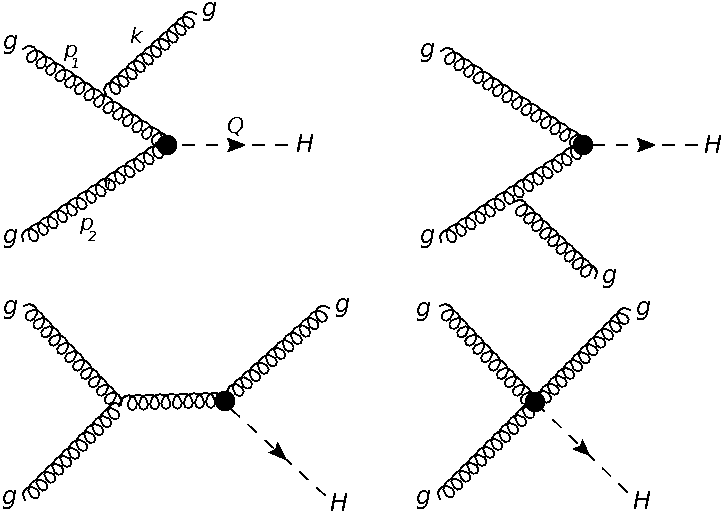}
  \caption{\sf
    The NLO Feynman diagrams for real-parton radiation 
    in the process of Higgs-boson production
    in gluon--gluon fusion: the $gg$ channel.  
    }
  \label{fig:ggHg}
\end{figure}
\begin{figure}[!ht]
  \centering
  \includegraphics[width=50mm]{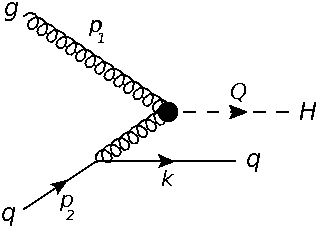}
  \caption{\sf
    The NLO Feynman diagram for real-parton radiation 
    in the process of Higgs-boson production
    in gluon--gluon fusion: the $gq$ channel.  
    }
  \label{fig:gqHq}
\end{figure}
\begin{figure}[!ht]
  \centering
  \includegraphics[width=60mm]{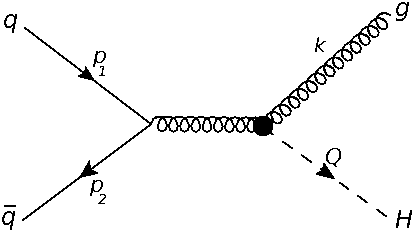}
  \caption{\sf
    The NLO Feynman diagram for real-parton radiation 
    in the process of Higgs-boson production
    in gluon--gluon fusion: the $q\bar{q}$ channel.  
    }
  \label{fig:qqHg}
\end{figure}

For the $gg$-channel NLO subprocess
\begin{equation}
g\; +\; g \:\longrightarrow\: H + g\,,
\label{eq:NLOproc-gg}
\end{equation}
shown in Fig.~\ref{fig:ggHg}, the matrix element squared reads
\begin{equation}
|{\cal M}_{gg}^{\nlo}|^2 = 8\pi\alpha_sC_A\,\frac{1}{zQ^2}\,
\frac{1+z^4+\alpha^4+\beta^4}{\alpha\beta}\,|{\cal M}_{gg}^{\lo}|^2.
\label{eq:NLOME-gg}
\end{equation}
For the $qg$-channel NLO subprocess
\begin{equation}
g\; +\; q \:\longrightarrow\: H + q\,,
\label{eq:NLOproc-gq}
\end{equation}
shown in Fig.~\ref{fig:gqHq},
one obtains
\begin{equation}
|{\cal M}_{gq}^{\nlo}|^2 = 8\pi\alpha_sC_F\,\frac{1}{zQ^2}\,
\frac{1+\beta^2}{\alpha}\,|{\cal M}_{gg}^{\lo}|^2,
\label{eq:NLOME-gq}
\end{equation}
Finally, for the $q\bar{q}$ channel
\begin{equation}
q\; +\; \bar{q} \:\longrightarrow\: H + g\,,
\label{eq:NLOproc-qq}
\end{equation}
see Fig.~\ref{fig:qqHg}, one has
\begin{equation}
|{\cal M}_{q\bar{q}}^{\nlo}|^2 = 8\pi\alpha_s  C_F\, \frac{8}{3}\,\frac{1}{zQ^2}\,
\left(\alpha^2+\beta^2\right)\,|{\cal M}_{gg}^{\lo}|^2.
\label{eq:NLOME-qq}
\end{equation}
This last process, unlike the previous ones, 
is not generated by the backward-evolution PSMC
starting from the $gg\rightarrow H$ hard process,
hence in the \krknlo{} method, its contribution cannot be
treated by NLO-reweighting of events generated by the main
branch of the LO PSMC algorithm.
It has to be added as an extra tree-level LO process to PSMC.
Moreover, it is free of collinear and soft singularities.
This poses no problem as most of present-day PSMCs implement such a process.

\subsection{CS dipoles and MC matrix elements}
In the following we shall elaborate mainly on the hadron--hadron collision
producing the Higgs boson or $Z/\gamma^*$ (Drell--Yan process).
However, components of the \krknlo{} method defined here will be
also applicable to any LO process $a+\bar{a} \rightarrow X$
and the corresponding  $ a+b \rightarrow X+c$
where $X=H, Z/\gamma, W^\pm, ZZ, W^+W^-$ or any other colour-neutral
heavy object; $a,b=q,\bar{q},g$ are initial coloured partons and $c$
is an additional parton emitted at the NLO level.

\begin{figure}[!h]
\centering
\includegraphics[width=0.90\textwidth]{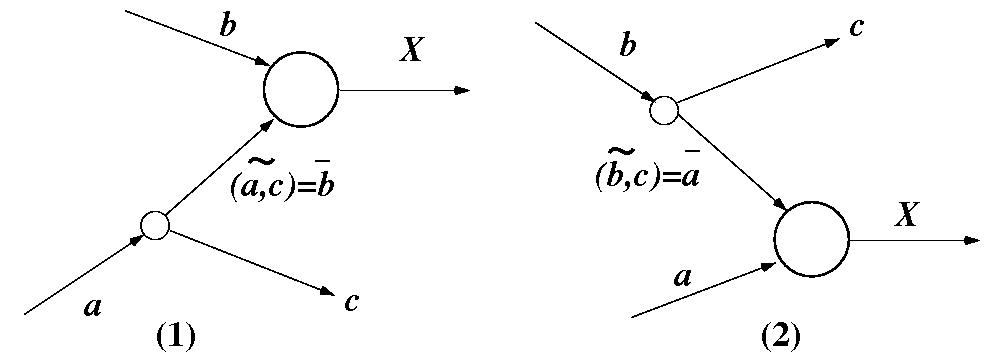}
\caption{\sf 
Kinematics of a single channel with one CS splitting.
}
\label{fig:CSdipole}
\end{figure}
In the following formulation of the \krknlo-method components,
the CS dipoles will serve us as useful auxiliary objects.
They  are formed by an initial-state (on-shell) 
emitter $a$ from one hadron and a spectator parton $b$ from another hadron%
\footnote{The role of the spectator is to provide for momentum and colour conservation.},
see Fig.~\ref{fig:CSdipole}.
Following closely the notation of the CS work~\cite{Catani:1996vz},
the emitter $a$ splits into an off-shell $\widetilde{ac}=\bar{b}$
entering into the hard process and an emitted parton $c$.
The CS dipoles $\Dmf^{(ac,b)}$ relevant for processes of our interest
are proportional to $\bar{P}_{ \widetilde{ac} ,a}$, 
the DGLAP kernel for the $ a\to \widetilde{ac}$ splitting%
\footnote{In the case of the emitted parton $c$ being the gluon one
gets $\widetilde{ac}\equiv a$.}.

\begin{table}[t]
\centering
\begin{tabular}{|c|c|c|c|}
\hline \hline
\multicolumn{2}{|c|}{$a+b\to c+H$}   &  
\multicolumn{2}{|c|}{$a+b \to c+Z/\gamma$}   \\
\hline 
  $(a,b)$       & $(ac,b)$            & $(a,b)$ & $(ac,b)$  \\
\hline 
  $(g,g)$       & $(gg,g)$            & $(q,\bar{q})$ & $(qg,\bar{q})$ \\
  $(q,g)$       & $(qq,g)$            & $(\bar{q},q)$ & $(\bar{q}g,q)$ \\
  $(\bar{q},g)$ & $(\bar{q}\bar{q},g)$& $(g,q)$       & $(gq,q)$ \\
                &                     & $(g,\bar{q})$ & $(g\bar{q},\bar{q})$ \\
\hline \hline                        
\end{tabular}
\caption{\sf
List of indices labelling the CS or MC soft-collinear counter terms for all the NLO channels 
(except the $q\bar{q}$ channel in Higgs production)
of annihilation processes.
Indices $(a,b)$ denote initial partons (channel),
while $(ac,b)$ are labelling the CS/MC counter terms, 
with $a$ being an emitter and $b$ a spectator.
}
\label{tab:indices}
\end{table}

For the processes  of the annihilation $a\bar{a} \to X$ at the LO level,
such as the Higgs production and the DY process,
in each NLO channel  $ab \to cX $
we must have  $ \widetilde{ac} =\bar{b}$ in the NLO splitting.
In other words, the NLO splitting in the annihilation
processes is fully determined by $a$ and $b$%
\footnote{This is, of course, not true for other processes.}.
The above rules are illustrated in Fig.~\ref{fig:CSdipole} 
and possible indices are listed in Table~\ref{tab:indices}
for the emission from the incoming line $a$%
\footnote{Rules for emissions from the second incoming line are analogous.}.

Let us first define explicitly the MC distributions (matrix elements) and the CS dipoles representing
the initial-state real-parton emissions 
for the Higgs production process in $d=4+2\veps$ dimensions:
\begin{enumerate}[(A)]
\item
For the $g + g \rightarrow H + g$ channel a typical/representative
distribution of PSMC, summing the emissions from both incoming gluons, is
\begin{equation}
|{\cal M}_{gg\rightarrow Hg}^{\mc}|^2 =
8\pi\alpha_s\,\mu^{-2\epsilon}\,\frac{1}{Q^2}\,
\frac{1}{\alpha\beta}\,(1-z)\hat{P}_{gg}(z;\epsilon) |{\cal M}_{gg\rightarrow H}^{\lo}|^2,
\label{eq:CSdip-ggH}
\end{equation}
where the $g\to g$ splitting function is given by 
\begin{equation}
\hat{P}_{gg}(z;\epsilon) = 2C_A\left[\frac{z}{1-z} + \frac{1-z}{z} + z(1-z)\right] =
C_A\frac{1 + z^4 + (1-z)^4}{z(1-z)}.
\label{eq:Pgg}
\end{equation}
It is equal to the sum of two CS dipoles
$ |{\cal M}_{gg\rightarrow Hg}^{\mc}|^2 = \Dmf^{(gg,g)} _{(1)}+  \Dmf^{(gg,g)}_{(2)}$, 
where
\begin{equation}
\Dmf^{(gg,g)}_{(1)}= \frac{\alpha}{\alpha+\beta} |{\cal M}_{gg\rightarrow Hg}^{\mc}|^2,
\qquad
\Dmf^{(gg,g)}_{(2)}= \frac{\beta}{\alpha+\beta} |{\cal M}_{gg\rightarrow Hg}^{\mc}|^2,
\label{eq:Dgg}
\end{equation}
with soft partition functions $ \frac{\alpha}{\alpha+\beta}$
and $\frac{\beta}{\alpha+\beta}$ separating the soft singularity evenly
between two incoming emitters.
Indices (1) and (2) are used to distinguish the above two dipoles.
\item
For the $g + q \rightarrow H + q$ channel we have 
(with a single soft-collinear pole the soft-partition functions are not needed):

\begin{equation}
|{\cal M}_{gq\rightarrow Hq}^{\mc}|^2 
= \Dmf^{(qq,g)}_{(1)}
=8\pi\alpha_s\,\mu^{-2\epsilon}\,\frac{1}{Q^2}\,
\frac{1}{\alpha}\,\hat{P}_{gq}(z;\epsilon) |{\cal M}_{gg\rightarrow H}^{\lo}|^2,
\label{eq:CSdip-gqH}
\end{equation}
where the $q\to g$ splitting function reads
\begin{equation}
\hat{P}_{gq}(z;\epsilon) = C_F\left[\frac{1 + (1-z)^2}{z} + \epsilon\, z\right] .
\label{eq:Pgq}
\end{equation}
\item
Finally, for the $g + \bar{q} \rightarrow H + \bar{q}$ channel, 
the CS dipole and MC distribution
is the same as the previous one for quarks.
\end{enumerate}
The above distributions agree with these 
used in the \powheg-method construction of Ref.~\cite{Hamilton:2009za}.

For the sake of completeness, let us collect the CS dipoles
and MC distributions 
already known from Refs.~\cite{Jadach:2011cr,Jadach:2015mza},
with the $q\to q$ and $g\to q$ splittings.
They will be needed in the following to define the transition matrix $K$
from the $\msbar$ to MC factorisation scheme
for all PDFs.

\begin{enumerate}[(A)]
\item
For the $q + \bar{q} \rightarrow Z + g$ channel, the MC distribution reads:
\begin{equation}
|{\cal M}_{q\bar{q}\rightarrow Zg}^{\mc}|^2 
= \Dmf^{(q g ,\bar{q})}_{(1)} +\Dmf^{(\bar{q} g ,q)}_{(2)}
= 8\pi\alpha_s\,\mu^{-2\epsilon}\,\frac{1}{Q^2}\,
\frac{1}{\alpha\beta}\,(1-z)\hat{P}_{qq}(z;\epsilon) 
   |{\cal M}_{q\bar{q}\rightarrow Z}^{\lo}|^2,
\label{eq:CSdip-qqZ}
\end{equation}
where 
\begin{equation}
\hat{P}_{qq}(z;\epsilon) = C_F\left[\frac{1+z^2}{1-z} + \epsilon (1-z) \right],
\label{eq:Pqq}
\end{equation}
and the soft-partition function is used again:
\begin{equation}
\Dmf^{(qg ,\bar{q})}_{(1)}
  = \frac{\alpha}{\alpha+\beta} |{\cal M}_{q\bar{q}\rightarrow Zg}^{\mc}|^2,
\qquad
\Dmf^{(\bar{q}g ,q)}_{(2)}
  = \frac{\beta}{\alpha+\beta} |{\cal M}_{q\bar{q}\rightarrow Zg}^{\mc}|^2.
\label{eq:DqqZ}
\end{equation}
\item 
For the $q + g \rightarrow Z + q$ channel we have
(the soft partition function in the MC distribution is not necessary):
\begin{equation}
|{\cal M}_{qg\rightarrow Zq}^{\mc}|^2 
= \Dmf^{(gq ,q)}_{(1)}
= 8\pi\alpha_s\,\mu^{-2\epsilon}\,\frac{1}{Q^2}\,
\frac{1}{\alpha}\,\hat{P}_{qg}(z;\epsilon) |{\cal M}_{q\bar{q}\rightarrow Z}^{\lo}|^2,
\label{eq:CSdip-qgZ}
\end{equation}
where 
\begin{equation}
\hat{P}_{qg}(z;\epsilon) = T_R\left[z^2 + (1-z)^2 + 2\epsilon\, z(1-z)\right] .
\label{eq:Pqg}
\end{equation}
\end{enumerate}
It should be stressed that all the above MC distributions and CS dipoles
are basically in the exclusive (unintegrated) form.

All the above relations between the MC distributions 
and the exclusive MC/CS counter terms
for any annihilation processes 
can be summarised in a compact formula as follows:
\begin{equation}
|{\cal M}_{ab\rightarrow cX}^{\mc}|^2 
= \Dmf^{(ac ,b)}_{(1)} +\Dmf^{(bc ,a)}_{(2)},
\label{eq:twodipoles}
\end{equation}
where translation from the indices $(ab)$ to $(abc)$ is unique
for a given annihilation process and for a given initial parton splitting,
as demonstrated explicitly in Table~\ref{tab:indices} for
the splitting of the initial parton $a$, see also Fig.~\ref{fig:CSdipole}.
Moreover, on the RHS of the above relation only one of $\Dmf$'s
is nonzero, except for the $c=g$ case (gluonstrahlung),
but in this case both $\Dmf$'s are equal.
Hence, there is in practice one-to-one correspondence 
$(ab) \leftrightarrow (abc)$ for all annihilation processes, 
to be often exploited in the following section.

\subsection{Integrated CS dipoles and counter terms of MC scheme}
For the purpose of installing virtual parts (using PDF momentum sum rules)
in the MC distributions (soft-collinear counter terms) and defining
the $K$-matrix for transforming PDFs from the $\msbar$ to MC scheme,
we need to integrate partly all distributions
defined in the previous subsection, keeping the $z=1-\alpha-\beta$ variable fixed.

A $z$-dependent differential cross section corresponding 
to the real-emission MC matrix elements can be expressed 
in the following way:
\begin{equation}
\frac{1}{z} \frac{d\hat{\sigma}_{ab,R}^{\mc}(z,\epsilon)}{dz} 
=  \frac{1}{2Q^2} \int |{\cal M}_{ab\rightarrow Xc}^{\mc}|^2 
   d\hat{\Phi} = \sigma_0\, \hat{\Gamma}^{\mc}_{ab,R}(z,\epsilon),
\label{eq:sigCS}
\end{equation}
where $\hat{\Gamma}^{\mc}_{ab,R}(z,\epsilon)$ is the MC real-emission function corresponding to the partly integrated 
MC distribution of the previous subsection for a given process: $a+b \rightarrow X+c$. 
The integration element $d\hat{\Phi}$ 
can be expressed in terms of the Sudakov variables as follows
\begin{equation}
d\hat{\Phi} = 
\frac{1}{8\pi} \left(\frac{4\pi}{s}\right)^{-\epsilon} 
\frac{1}{\Gamma(1+\epsilon)}(\alpha\beta)^{\epsilon}
 \delta(1 - z - \alpha - \beta) 
 \theta(\alpha)\theta(1-\alpha)
 \theta(\beta)\theta(1-\beta)
 \theta(1-\alpha-\beta)\;
 d\alpha\, d\beta\,.
\label{eq:dPhi}
\end{equation}
The above expressions are defined in $d = 4 + 2\epsilon$ 
dimensions in order to regularise, in the usual way, the soft and collinear
singularities of the real-parton radiation.

Using the {\em exact} NLO matrix element,
one can similarly write, for each channel $ab$,
a regularised partly integrated
NLO cross section for real-parton emission:
\begin{equation}
\frac{1}{z} \frac{d\hat{\sigma}_{ab,R}^{\nlo}(z,\epsilon)}{dz} 
=  \frac{1}{2Q^2} \int |{\cal M}_{ab\rightarrow Xc,R}^{\nlo}|^2 
d\hat{\Phi} = \sigma_0 \hat{\rho}^{\nlo}_{ab,R}(z,\epsilon).
\label{eq:sigNLO}
\end{equation}

Following the relation of Eq.~(\ref{eq:twodipoles}), one may also
define the relation of the integrated MC distribution to
the individual integrated soft-collinear counter terms:
\begin{equation}
 \hat{\Gamma}^{\mc}_{ab,R}(z,\epsilon)
=\hat{\Lambda}^{\mc}_{(\widetilde{ac},b),R}(z,\epsilon)
+\hat{\Lambda}^{\mc}_{(\widetilde{bc},a),R}(z,\epsilon)
\end{equation}
where $\hat{\Lambda}_R$
are the corresponding integrals $\int d\hat{\Phi}\; \Dmf$
as in Eq.~(\ref{eq:sigCS}).
However, contrary to the CS counter terms,
the counter terms $\hat{\Lambda}^{\mc}$ of the MC scheme
(and the $\hat{\Gamma}^{\mc}$  radiation functions as well)
will also include virtual corrections,
calculated using the momentum sum rules, see next subsections for details.

Let us calculate all the above objects with more details for
the $gg\to Hg$ channel and then, 
skipping details of analytical integration, for other channels.

\subsection{$gg \rightarrow Hg$ channel}
\label{subsec:CFgg}
A real-emission part of the MC radiation function
results from the following integration%
\footnote{We employ here and in the following
   a shorthand notation $\delta_{x=y}\equiv \delta(x-y)$.}
\begin{equation}
\begin{split}
&
\hat{\Gamma}^{\mc}_{gg,R}(z,\epsilon) 
=\frac{1}{\sigma_0} \frac{1}{2Q^2}  \int |{\cal M}_{gg\rightarrow Hg}^{\mc}|^2 d\hat{\Phi} 
= \frac{2C_A \alpha_s}{2\pi}\, 
  \left(\frac{4\pi\mu^2}{s}\right)^{-\epsilon} \frac{1}{\Gamma(1+\epsilon)}
\\&~~~~~~~\times
  \left[z + \frac{(1-z)^2}{z} + z(1-z)^2\right] 
  \int_0^1 d\alpha \int_0^1 d\beta \, (\alpha\beta)^{-1+\epsilon}\,
  \delta_{1 - z = \alpha + \beta}
\\&
= \frac{\alpha_s}{2\pi}\,2C_A \left(\frac{4\pi\mu^2}{Q^2}\right)^{-\epsilon} 
\frac{\Gamma(1+\epsilon)}{\Gamma(1+2\epsilon)}\,
z^{-\epsilon}\,(1-z)^{-1+2\epsilon}\; 
\frac{2}{\epsilon}\left[z + \frac{(1-z)^2}{z} + z(1-z)^2\right].
\label{eq:Gamgg4}
\end{split}
\end{equation}
Using the standard expansion
$ (1-z)^{-1+2\epsilon} = 
    \frac{1}{2\epsilon}\delta(1-z)
  + \big(\frac{1}{1-z}\big)_+ 
  + 2\epsilon\big(\frac{\ln(1-z)}{1-z}\big)_+ $
we obtain
\begin{equation}
\begin{split}
&\hat{\Gamma}^{\mc}_{gg,R}(z,\epsilon) 
= \frac{2C_A \alpha_s}{2\pi}
\left(\frac{4\pi\mu^2}{Q^2}\right)^{-\epsilon}
\frac{\Gamma(1+\epsilon)}{\Gamma(1+2\epsilon)}
\Bigg\{
\frac{\delta(1-z)}{\epsilon^2} 
+ \frac{2}{\epsilon} \left[ \left(\frac{1}{1-z}\right)_+ + \frac{1}{z} - 2 + z(1-z) \right]
\\ &
+  4\left[\frac{1}{z}\left(\frac{\ln(1-z)}{1-z}\right)_+ - [2 - z(1-z)]\ln(1-z)\right]
- 2\left[\left(\frac{1}{1-z}\right)_+ + \frac{1}{z} - 2 + z(1-z)\right]\ln z 
\Bigg\}.
\end{split}
\label{eq:GamggR}
\end{equation}
The NLO real correction according to Ref.~\cite{Djouadi:1991tka} reads
\begin{equation}
\begin{split}
& \hat{\rho}^{\nlo}_{gg,R}(z,\epsilon) 
 = \frac{2C_A \alpha_s}{2\pi}
  \left(\frac{4\pi\mu^2}{Q^2}\right)^{-\epsilon}  
  \frac{\Gamma(1+\epsilon)}{\Gamma(1+2\epsilon)}\,
  \Bigg\{ \frac{\delta(1-z)}{\epsilon^2} 
+ \frac{2}{\epsilon} \left[ \left(\frac{1}{1-z}\right)_+ + \frac{1}{z} - 2 + z(1-z) \right]
\\ &~~~~~~~~
+  4\left[\frac{1}{z}\left(\frac{\ln(1-z)}{1-z}\right)_+ - [2 - z(1-z)]\ln(1-z)\right]
\\ &~~~~~~~~
-  2\left[\left(\frac{1}{1-z}\right)_+ + \frac{1}{z} - 2 + z(1-z)\right]\ln z 
- \frac{11}{6}\frac{(1-z)^3}{z}
\Bigg\}.
\end{split}
\label{eq:rhoggNLO}
\end{equation}
From the above equations we readily obtain 
the NLO real coefficient function in the MC scheme:
\begin{equation}
 H_{gg,R}^{\rm MC}(z) = \hat{\rho}^{\nlo}_{gg,R}(z,\epsilon) - \hat{\Gamma}^{\mc}_{gg,R}(z,\epsilon)
 = \frac{\alpha_s}{2\pi}\, 2C_A \left\{- \frac{11}{6}\frac{(1-z)^3}{z} \right\}.
 \label{eq:C2RMC1}
\end{equation}
The same expression is obtained in $4$ dimensions by means 
of performing first the MC-dipole
subtraction and then integrating the finite result over the phase space:
\begin{equation}
\begin{split}
& H_{gg,R}^{\rm MC}(z) =
 \frac{1}{\sigma_0} \frac{1}{2Q^2}  
 \int \left[|{\cal M}_{gg}^{\nlo}|^2  - |{\cal M}_{gg\rightarrow Hg}^{\mc}|^2 \right] d\Phi
 =\frac{ 2C_A \alpha_s}{2\pi}
  \frac{1}{2z} 
\\&~~~~~~\times
  \int_0^1d\alpha \int_0^1 d\beta\, \delta_{1-z=\alpha+\beta}
  \frac{1 + z^4 + \alpha^4 + \beta^4 - 2[z^2 + (1-z)^2 + z^2(1-z)^2]}{\alpha\beta}
\\&
 =  \frac{\alpha_s}{2\pi}\, 2C_A \left\{- \frac{11}{6}\frac{(1-z)^3}{z} \right\}.
\end{split}
\label{eq:C2RMC2}
\end{equation}

A virtual correction to the above MC radiation function $\hat{\Gamma}_{gg}$
is calculated from the momentum sum rules:
\begin{equation}
\hat{\Gamma}^{\mc}_{gg,V}(z,\epsilon) = - \delta(1-z) \int_0^1 dz\, z 
 \left[ \hat{\Gamma}^{\mc}_{gg,R}(z,\epsilon) 
 + 2n_f\cdot 2\hat{\Gamma}^{\mc}_{qg}(z,\epsilon)
 \right],
\label{eq:GamVMC}
\end{equation}
where $n_f$ is the number of fermions.
The first part in the above virtual correction 
resulting from integration over the first term in brackets 
on RHS reads as follows:
\begin{equation}
\begin{split}
& \hat{\Gamma}^{\mc}_{gg,V_1}(z,\epsilon) 
=  - \delta(1-z) \int_0^1 dz\, z\,  \hat{\Gamma}^{\mc}_{gg,R}(z,\epsilon) 
\\&~~~~~
= \delta(1-z) \, \frac{\alpha_s}{2\pi}\,2C_A \,
  \left(\frac{4\pi\mu^2}{Q^2}\right)^{-\epsilon} 
  \frac{\Gamma(1+\epsilon)}{\Gamma(1+2\epsilon)}\,
  \left\{-\frac{1}{\epsilon^2} + \frac{11}{6}\frac{1}{\epsilon} 
       - \frac{341}{72} - \frac{\pi^2}{3} \right\}.
\end{split}
\label{eq:GamV1ggMC}
\end{equation}

In order to calculate the second part
to the virtual correction in RHS of Eq.~(\ref{eq:GamVMC})
we need to know first the following 
MC radiation functions for the $g \rightarrow q$ transition, 
e.g.\ from the process $q + g \rightarrow Z + q$:
\begin{equation} 
\begin{split} 
&\hat{\Gamma}^{\mc}_{q g}(z,\epsilon)
= \frac{1}{\sigma_0} \frac{1}{2Q^2}  
   \int |{\cal M}_{qg\rightarrow Zq}^{\mc}|^2  d\hat{\Phi} 
=\frac{\alpha_s}{2\pi}\,T_R \left(\frac{4\pi\mu^2}{Q^2}\right)^{-\epsilon} 
\frac{\Gamma(1+\epsilon)}{\Gamma(1+2\epsilon)}\,
\\&~~~~~\times
\Bigg\{ \frac{1}{\epsilon}\left[z^2 + (1-z)^2\right]  
+ \left[z^2 + (1-z)^2\right]\ln \frac{(1-z)^2}{z} + 2z(1-z)
+ {\cal O}(\veps)
\Bigg\},
\label{eq:Gamqg5}
\end{split} 
\end{equation} 
where $|{\cal M}_{qg\rightarrow Zq}^{\mc}|^2$ is shown in Eq.~(\ref{eq:CSdip-qgZ}).

Using the above result we can cross-check 
the formula for the gluon-channel MC radiation function 
of the DY process calculated previously in $4$ dimensions
in Ref.~\cite{Jadach:2015mza}.
For the exact NLO contribution Ref.~\cite{Altarelli:1979ub} provides
\begin{equation}
\begin{split}
&
\hat{\rho}^{\nlo}_{qg\rightarrow Zq}(z,\epsilon) 
= \frac{\alpha_s}{2\pi}\,T_R \, 
  \left(\frac{4\pi\mu^2}{Q^2}\right)^{-\epsilon} 
  \frac{\Gamma(1+\epsilon)}{\Gamma(1+2\epsilon)}\,
\\&~~~~~~~\times
  \Bigg\{ \frac{1}{\epsilon}\left[z^2 + (1-z)^2\right]  
 + \left[z^2 + (1-z)^2\right]\ln \frac{(1-z)^2}{z} 
 - \frac{7}{2}z^2 + 3z + \frac{1}{2}\Bigg\}.
\end{split}
\label{eq:rhoqgNLO}
\end{equation}
Then, the resulting coefficient function for the DY process in the MC scheme reads
\begin{equation}
 C_{qg}^{\rm MC}(z) 
 =  \hat{\rho}^{\nlo}_{qg\rightarrow Zq}(z,\epsilon) 
  - \hat{\Gamma}^{\mc}_{qg}(z,\epsilon)
 = \frac{\alpha_s}{2\pi}\, T_R \left\{\frac{1}{2}(1-z)(1+3z) \right\},
 \label{eq:C2qgMC}
\end{equation}
 which agrees with our previous result, given in Ref.~\cite{Jadach:2015mza}.

For the sake of completeness, 
the corresponding coefficient function in the $\overline{\rm MS}$ 
factorisation scheme reads:
\begin{equation}
C_{qg}^{\overline{\rm MS}}(z) 
= \frac{\alpha_s}{2\pi}\,T_R \,
   \left\{ \left[z^2 + (1-z)^2\right]\ln \frac{(1-z)^2}{z} 
 - \frac{7}{2}z^2 + 3z + \frac{1}{2}\right\}
\label{eq:C2qgMSbar}
\end{equation}
and the transition matrix element transforming part of
gluon $\msbar$ PDF into the quark PDF in the MC scheme is given by
\begin{equation}
K^{\rm MC}_{qg}(z) = C_{qg}^{\overline{\rm MS}}(z) - C_{qg}^{\rm MC}(z) 
= \frac{\alpha_s}{2\pi}\,T_R \,
 \left\{ \left[z^2 + (1-z)^2\right]\ln \frac{(1-z)^2}{z}  + 2z(1-z) \right\}.
\label{eq:DeltaC2qg}
\end{equation}

The above was obtained by comparing
the NLO coefficient functions for the DY process
in the  $\overline{\rm MS}$ and MC schemes.
However, exactly the same result can be
obtained alternatively from the difference of 
the soft-collinear counter terms in these two schemes:
\begin{equation}
K^{\rm MC}_{qg}(z) = \left[ \hat{\Lambda}_{qg}^{\rm MC}(z,\epsilon) 
         - \hat{\Lambda}_{qg}^{\overline{\rm MS}}(z,\epsilon) \right]_{\epsilon = 0},
\label{eq:DeltaC2qgCC}
\end{equation}
where the universal MC-scheme counter term corresponding to the $g\to q$ transition is given by
\begin{equation}
\hat{\Lambda}_{qg}^{\rm MC}(z,\epsilon) = \hat{\Gamma}_{qg}^{\rm MC}(z,\epsilon),
\label{eq:CCqgMC}
\end{equation}
where $\hat{\Gamma}_{qg}^{\rm MC}(z,\epsilon)$ is defined in Eq.~(\ref{eq:Gamqg5}),
and the $\msbar$ counter term is
\begin{equation}
\hat{\Lambda}_{qg}^{\overline{\rm MS}}(z,\epsilon)
=\frac{\alpha_s}{2\pi}\,T_R \left(\frac{4\pi\mu^2}{Q^2}\right)^{-\epsilon} 
\frac{\Gamma(1+\epsilon)}{\Gamma(1+2\epsilon)}\,
\frac{1}{\epsilon}\left[z^2 + (1-z)^2\right].
\label{eq:GamqgMSbar}
\end{equation}

After this brief detour to the DY process, we can now complete the calculation of the virtual correction
to the MC radiation function for the $gg\to Hg$ channel.
Using Eq.~(\ref{eq:Gamqg5}), the second term in RHS 
of Eq.~(\ref{eq:GamVMC}) is calculated:
\begin{equation}
\begin{split}
\hat{\Gamma}^{\mc}_{gg,V_2}(z,\epsilon) 
&=  - \delta(1-z)\cdot 4n_f \int_0^1 dz\, z\; \hat{\Gamma}^{\mc}_{qg}(z,\epsilon)  
\\&
= \delta(1-z) \, \frac{\alpha_s}{2\pi}\, n_f T_R\, 
  \left(\frac{4\pi\mu^2}{Q^2}\right)^{-\epsilon} 
  \frac{\Gamma(1+\epsilon)}{\Gamma(1+2\epsilon)}\,
  \left\{-\frac{4}{3}\frac{1}{\epsilon} + \frac{59}{18} \right\}.
\end{split}
\label{eq:GamV2ggMC}
\end{equation} 
The complete result for virtual correction to the $gg\to Hg$ MC radiation function,
obtained from the momentum sum rule of Eq.~(\ref{eq:GamVMC}),
reads as follows:
\begin{equation}
\begin{split}
&\hat{\Gamma}^{\mc}_{gg,V}(z,\epsilon) =
 \hat{\Gamma}^{\mc}_{gg,V_1}(z,\epsilon) + \hat{\Gamma}^{\mc}_{gg,V_2}(z,\epsilon) 
= \delta(1-z) \, \frac{2C_A \alpha_s}{2\pi}
\left(\frac{4\pi\mu^2}{Q^2}\right)^{-\epsilon} 
\frac{\Gamma(1+\epsilon)}{\Gamma(1+2\epsilon)}\,
\\&~~~~~~~~~~~~~~~~~~~~~\times
\Bigg\{-\frac{1}{\epsilon^2} + \frac{1}{\epsilon}\frac{11-4T_f/C_A}{6} 
- \frac{341}{72} - \frac{\pi^2}{3} + \frac{T_f}{C_A}\frac{59}{36}\Bigg\},
\end{split}
\label{eq:GamVggMC}
\end{equation}
where $T_f = n_f T_R$. 

The complete MC radiation function for $gg\to Hg$ process
is obtained finally in the following explicit form:
\begin{equation}
\begin{split}
\hat{\Gamma}^{\mc}_{gg}(z,\epsilon) 
&= \frac{\alpha_s}{2\pi}\,2C_A \,
  \left(\frac{4\pi\mu^2}{Q^2}\right)^{-\epsilon}\, 
  \frac{\Gamma(1+\epsilon)}{\Gamma(1+2\epsilon)}\,
\\& \times 
\Bigg\{
\frac{2}{\epsilon} 
  \left[ 
    \delta(1-z) \frac{11-4T_f/C_A}{12} 
     + \left(\frac{1}{1-z}\right)_+ + \frac{1}{z} - 2 + z(1-z)  
  \right]
\\ &~~~~~
 -\delta(1-z)
   \left[ \frac{\pi^2}{3} + \frac{341}{72} - \frac{T_f}{C_A}\frac{59}{36} \right] 
\\ &~~~~~
  +  4\left[\frac{\ln(1-z)}{1-z}\right]_+ + 2\left[\frac{1}{z} - 2 
          + z(1-z) \right]\ln\frac{(1-z)^2}{z}  - 2 \frac{\ln z}{1-z} 
\Bigg\}.
\end{split}
\label{eq:Gamgg}
\end{equation}

Let us also calculate  the coefficient function
in the MC scheme for the $gg\to Hg$ channel.
Using the exact NLO virtual correction of Ref.~\cite{Djouadi:1991tka}:
\begin{equation}
\begin{split}
&
\hat{\rho}^{\nlo}_{gg,V}(z,\epsilon) 
 =  \delta(1-z) \, \frac{\alpha_s}{2\pi}\,2C_A \,
    \left(\frac{4\pi\mu^2}{Q^2}\right)^{-\epsilon} 
\\&
    \frac{\Gamma(1+\epsilon)}{\Gamma(1+2\epsilon)}\,
    \Bigg\{-\frac{1}{\epsilon^2} + \frac{1}{\epsilon}\frac{11-4T_f/C_A}{6} 
    +\frac{11}{6} + \frac{\pi^2}{3}  - \frac{11-4T_f/C_A}{6} \ln\frac{Q^2}{\mu^2} 
\Bigg\}.
\end{split}
\label{eq:rhoVggNLO}
\end{equation}
we obtain the following virtual part of the coefficient function 
in the MC scheme (with the usual $\mu^2 = Q^2$ assignment):
\begin{equation}
 H_{gg,V}^{\rm MC}(z) 
 = \hat{\rho}^{\nlo}_{gg,V}(z,\epsilon) - \hat{\Gamma}^{\mc}_{gg,V}(z,\epsilon)
 = \frac{\alpha_s}{2\pi}\, 2C_A \,\delta(1-z)\left[ \frac{473}{72} + \frac{2\pi^2}{3} 
   - \frac{T_f}{C_A}\frac{59}{36}\right] .
\label{eq:C2VMC}
\end{equation}
Combining the real and virtual contributions
of Eqs.~(\ref{eq:C2RMC1}) and (\ref{eq:C2VMC}),
the NLO coefficient function for the $gg\to Hg$ process in the MC factorisation  scheme reads:
\begin{equation}
 H_{gg}^{\rm MC}(z) 
 = \frac{\alpha_s}{2\pi}\, 2C_A \left\{ \delta(1-z) \left(\frac{2}{3}\pi^2 + \frac{473}{72}
  - \frac{59}{36} \frac{T_f}{C_A}\right) - \frac{11}{6}\frac{(1-z)^3}{z} \right\}.
 \label{eq:C2ggMC}
\end{equation}

The analogous coefficient function in the $\msbar$ factorisation scheme 
is obtained from Eqs.~(\ref{eq:rhoggNLO}) and (\ref{eq:rhoVggNLO}), 
after the standard subtraction of the  $\msbar$ soft-collinear counter terms
(with $\mu^2 = Q^2$) reads as follows:
\begin{equation}
\begin{split}
H_{gg}^{\overline{\rm MS}}(z) 
= \frac{\alpha_s}{2\pi}\, 2C_A
&
   \Bigg\{ \delta(1-z) \left(\frac{\pi^2}{3} + \frac{11}{6}\right) 
+ 4\left[\frac{\ln(1-z)}{1-z} \right]_+
\\ & 
+ 2\left[\frac{1}{z} - 2 + z(1-z) \right]\ln\frac{(1-z)^2}{z}  
- 2 \frac{\ln z}{1-z} - \frac{11}{6}\frac{(1-z)^3}{z}\Bigg\} .
\end{split}
\label{eq:C2ggMSbar}
\end{equation}

With all the above results at hand we are also ready to determine the element
$g \to g$ of the transition matrix for transforming PDFs from the $\msbar$
to MC scheme:
\begin{equation}
\begin{split}
&
K^{\rm MC}_{gg}(z) 
=  \frac{1}{2} \left[ H_{gg}^{\overline{\rm MS}}(z)\; -\; H_{gg}^{\rm MC}(z) \right] 
=   \frac{\alpha_s}{2\pi}\, C_A 
 \Bigg\{
  - \delta(1-z)\left( \frac{\pi^2}{3} + \frac{341}{72} - \frac{59}{36}\frac{T_f}{C_A}\right) 
\\&~~~~~~~~~~~~~~~~~~~~~~~
    4\left[\frac{\ln(1-z)}{1-z} \right]_+ 
  + 2\left[\frac{1}{z} - 2 + z(1-z) \right]\ln\frac{(1-z)^2}{z}  - 2 \frac{\ln z}{1-z} 
 \Bigg\}.
\end{split}
\label{eq:DeltaC2gg}
\end{equation}

The same $K^{\rm MC}_{gg}(z)$
it can be also obtained from the difference of the collinear counter terms:
\begin{equation}
K^{\rm MC}_{gg}(z) =
\left[  \hat{\Lambda}_{gg}^{\rm MC}(z,\epsilon) 
      - \hat{\Lambda}_{gg}^{\overline{\rm MS}}(z,\epsilon) \right]_{\epsilon = 0},
\label{eq:DeltaC2ggCC}
\end{equation}
where the universal MC counter term $\hat{\Lambda}_{gg}^{\rm MC}(z,\epsilon)$ corresponding to the $g\to g$ transition 
can be expressed in terms of the MC radiation function of Eq.~(\ref{eq:Gamgg})
as follows
\begin{equation}
\hat{\Lambda}_{gg}^{\rm MC}(z,\epsilon)  = \frac{1}{2} \hat{\Gamma}_{gg}^{\rm MC}(z,\epsilon) ,
\label{eq:CCggMC}
\end{equation}
and 
\begin{equation}
\begin{split}
&\hat{\Lambda}^{\overline{\rm MS}}_{gg}(z,\epsilon) 
= \frac{\alpha_s}{2\pi}\,2C_A \,
  \left(\frac{4\pi\mu^2}{Q^2}\right)^{-\epsilon}\, 
  \frac{\Gamma(1+\epsilon)}{\Gamma(1+2\epsilon)}\,
\\&~~~~~~~~~~~~~\times
  \frac{1}{\epsilon} 
\Bigg[ 
  \delta(1-z) \frac{11-4T_f/C_A}{12} 
+ \left(\frac{1}{1-z}\right)_+ + \frac{1}{z} - 2 + z(1-z)  
\Bigg]
\end{split}
\label{eq:GamggMSbar}
\end{equation}
is the corresponding $\msbar$ counter term.

\subsection{$gq \rightarrow Hq$ channel }
\label{subsec:CFgq}

The channel $g + q \rightarrow H + q$ is easier because
only real correction contributes at NLO.
The corresponding MC radiation function can be readily  obtained from the integral
\begin{equation}
\begin{split}
&
\hat{\Gamma}^{\mc}_{gq}(z,\epsilon)
=\frac{1}{\sigma_0} \frac{1}{2Q^2}  \int |{\cal M}_{gq\rightarrow Hq}^{\mc}|^2 d\hat{\Phi} 
\\&
=\frac{\alpha_s}{2\pi}C_F \left(\frac{4\pi\mu^2}{Q^2}\right)^{-\epsilon} 
\frac{\Gamma(1+\epsilon)}{\Gamma(1+2\epsilon)}
\left\{\frac{1 + (1-z)^2}{z}  \left[ \frac{1}{\epsilon} 
       +\ln\frac{(1-z)^2}{z}\right] + z 
\right\}.
\label{eq:Gamgq5}
\end{split}
\end{equation} 
The exact NLO correction taken from Ref.~\cite{Djouadi:1991tka} reads
\begin{equation}
\begin{split}
 \hat{\rho}^{\nlo}_{gq}(z,\epsilon) 
=& \frac{\alpha_s}{2\pi}\,C_F \,
  \left(\frac{4\pi\mu^2}{Q^2}\right)^{-\epsilon}  
  \frac{\Gamma(1+\epsilon)}{\Gamma(1+2\epsilon)}\,
\\ &~~~\times
  \Bigg\{
    \frac{1 + (1-z)^2}{z}  \left[ \frac{1}{\epsilon} +  \ln\frac{(1-z)^2}{z}\right] 
  - \frac{z^2 - 6z + 3}{2z} 
\Bigg\}.
\end{split}
\label{eq:rhogqNLO}
\end{equation}
Combining the above two functions,
the coefficient function for $gq\to Hq$ process
in the MC factorisation scheme reads
\begin{equation}
 H_{gq}^{\rm MC}(z) = \hat{\rho}^{\nlo}_{gq}(z,\epsilon) - \hat{\Gamma}^{\mc}_{gq}(z,\epsilon)
 = \frac{\alpha_s}{2\pi}\, C_F \left\{ - \frac{3}{2}\frac{(1-z)^2}{z} \right\}.
 \label{eq:C2gqMC}
\end{equation}
Exactly the same result can be obtained also
from the following integral in $4$ dimensions:
\begin{equation}
\begin{split}
H_{gq}^{\rm MC}(z) 
&= \frac{1}{\sigma_0} \frac{1}{2Q^2}  \int 
  \left[|{\cal M}_{gq}^{\nlo}|^2  
  - |{\cal M}_{gq\rightarrow Hq}^{\mc}|^2 
  \right] d\Phi
 \\&
 = \frac{\alpha_s}{2\pi}\, C_F\, 
   \frac{1}{z} \int_0^1d\alpha \int_0^1 d\beta\, \delta(1-z-\alpha-\beta)
   \frac{1 + \beta^2 - [1 + (1-z)^2]}{\alpha}
 \\&
 = \frac{\alpha_s}{2\pi}\, C_F \left\{ - \frac{3}{2}\frac{(1-z)^2}{z} \right\}. 
\label{eq:C2gqMC2}
\end{split}
\end{equation}

On the other hand, 
in the $\overline{\rm MS}$ factorisation scheme (keeping $\mu^2 = Q^2$),
from Eq.~(\ref{eq:rhogqNLO}) we can obtain (after the standard subtraction) 
the following coefficient function:
\begin{equation}
H_{gq}^{\overline{\rm MS}}(z) = 
\frac{\alpha_s}{2\pi} \, C_F 
\left\{ \frac{1 + (1-z)^2}{z} \ln\frac{(1-z)^2}{z}  - \frac{z^2 - 6z +3}{2z} \right\} .
\label{eq:C2gqMSbar}
\end{equation}

At this point we are able to define another element
of the matrix transforming the $\msbar$ gluon PDF into  the gluon PDF of the MC-scheme
\begin{equation}
K^{\rm MC}_{gq}(z) = H_{gq}^{\overline{\rm MS}}(z)\; -\; H_{gq}^{\rm MC}(z) 
= \frac{\alpha_s}{2\pi}\, C_F \left\{ \frac{1 + (1-z)^2}{z}\ln\frac{(1-z)^2}{z} + z\right\} .
\label{eq:DeltaC2gq}
\end{equation}

Alternatively, the same $K^{\rm MC}_{gq}(z)$
can also be obtained from as a difference of 
the soft-collinear counter terms in the MC  and $\msbar$ schemes:
\begin{equation}
K^{\rm MC}_{gq}(z) = 
\left[  \hat{\Lambda}_{gq}^{\rm MC}(z,\epsilon) 
      - \hat{\Lambda}_{gq}^{\overline{\rm MS}}(z,\epsilon) 
\right]_{\epsilon = 0},
\label{eq:DeltaC2gqCC}
\end{equation}
where, again, the universal MC-scheme counter term $ \hat{\Lambda}_{gq}^{\rm MC}(z,\epsilon)$ 
corresponding to the $q\to g$ transition
can be related to the MC radiation function $\hat{\Gamma}_{gq}^{\rm MC}(z,\epsilon)$ 
of Eq.~(\ref{eq:Gamgq5}) as follows:
\begin{equation}
\hat{\Lambda}_{gq}^{\rm MC}(z,\epsilon) = \hat{\Gamma}_{gq}^{\rm MC}(z,\epsilon),
\label{eq:CCgqMC}
\end{equation}
and 
\begin{equation}
\hat{\Lambda}_{gq}^{\overline{\rm MS}}(z,\epsilon) =
\frac{\alpha_s}{2\pi}C_F \left(\frac{4\pi\mu^2}{Q^2}\right)^{-\epsilon} 
\frac{\Gamma(1+\epsilon)}{\Gamma(1+2\epsilon)}
\frac{1}{\epsilon} \,\frac{1 + (1-z)^2}{z} 
\label{eq:GamgqMSbar}
\end{equation}
is the corresponding counter term in the $\msbar$ scheme. 

\subsection{Revisiting $q\bar{q} \rightarrow Zg$ channel}
\label{subsec:CFqqZ}

In Ref.~\cite{Jadach:2015mza} the virtual correction to the MC
counter term in the $q\bar{q}$ channel was calculated 
from the quark-number conservation sum rule 
(as minus the integral over $z$ of the real correction). 
This was justified for the DY process,
for which the gluon PDF did not get corrected at NLO
from the $\msbar$ to MC factorisation scheme.
Now, since we deal with the complete set of
parton--parton transitions, including the transformation/correction of the gluon PDF,
we have to rely on the momentum sum rule. 
For the pertinent channel this amounts to
\begin{equation}
\hat{\Gamma}^{\mc}_{q\bar{q},V}(z,\epsilon) = - \delta(1-z) \int_0^1 dz\, z 
\left[ \hat{\Gamma}^{\mc}_{q\bar{q},R}(z,\epsilon) 
       + \hat{\Gamma}^{\mc}_{gq}(z,\epsilon)
       + \hat{\Gamma}^{\mc}_{g\bar{q}}(z,\epsilon)
\right].
\label{eq:GamVqqMC}
\end{equation}

Using the formula for the MC real-radiation function from Appendix~B of Ref.~\cite{Jadach:2015mza}:
\begin{equation}
\begin{split}
\hat{\Gamma}^{\mc}_{q\bar{q},R}(z,\epsilon) =
\frac{\alpha_s}{2\pi}\,C_F \, \left(\frac{4\pi\mu^2}{Q^2}\right)^{-\epsilon} 
\frac{\Gamma(1+\epsilon)}{\Gamma(1+2\epsilon)}\,
\Bigg\{
\frac{2}{\epsilon^2}\,\delta(1-z) + \frac{2}{\epsilon}\frac{1+z^2}{(1-z)_+}  \qquad &
\\
+ \; 4(1+z^2)\left[\frac{\ln(1-z)}{1-z}\right]_+ - 2\frac{1+z^2}{1-z}\ln z + 2(1-z) 
&
\Bigg\},
\end{split}
\label{eq:GamRqqMC}
\end{equation} 
we can calculate the first part of the above virtual correction as follows:
\begin{equation}
\begin{split}
\hat{\Gamma}^{\mc}_{q\bar{q},V_1}(z,\epsilon) 
= & - \delta(1-z) \int_0^1 dz\, z\, \hat{\Gamma}^{\mc}_{q\bar{q},R}(z,\epsilon)  
\\
= & - \delta(1-z) \, \frac{\alpha_s}{2\pi}\,C_F \, 
  \left(\frac{4\pi\mu^2}{Q^2}\right)^{-\epsilon} 
\frac{\Gamma(1+\epsilon)}{\Gamma(1+2\epsilon)}\,
\int_0^1 dz\,z \,
\Bigg\{
\frac{2}{\epsilon^2}\,\delta(1-z) 
\\ 
& + \frac{2}{\epsilon}\frac{1+z^2}{(1-z)_+}  
  + 4(1+z^2)\left[\frac{\ln(1-z)}{1-z}\right]_+ 
  - 2\frac{1+z^2}{1-z}\ln z + 2(1-z)
\Bigg\}
\\
= & \,\delta(1-z)  \, \frac{\alpha_s}{2\pi}\,C_F \, 
\left(\frac{4\pi\mu^2}{Q^2}\right)^{-\epsilon} 
\frac{\Gamma(1+\epsilon)}{\Gamma(1+2\epsilon)}\,
\left\{-\frac{2}{\epsilon^2} 
+ \frac{17}{3}\frac{1}{\epsilon} - \frac{163}{18} - \frac{2\pi^2}{3} \right\}.
\end{split}
\label{eq:GamV1qqMC}
\end{equation} 
For the second part, using Eq.~(\ref{eq:Gamgq5}), we obtain
\begin{equation}
\begin{split}
\hat{\Gamma}^{\mc}_{q\bar{q},V_2}(z,\epsilon) 
= & - \delta(1-z)  \int_0^1 dz\, z\,  \left[ \hat{\Gamma}^{\mc}_{gq}(z,\epsilon)  
+   \hat{\Gamma}^{\mc}_{g\bar{q}}(z,\epsilon)  \right]
\\
= & - 2\,\delta(1-z) \, \frac{\alpha_s}{2\pi}\,C_F \, 
 \left(\frac{4\pi\mu^2}{Q^2}\right)^{-\epsilon} 
\frac{\Gamma(1+\epsilon)}{\Gamma(1+2\epsilon)}\,
\\ & \times \int_0^1 dz\,
\left\{\left[1 + (1-z)^2\right] 
\left[ \frac{1}{\epsilon} 
   +  \ln\frac{(1-z)^2}{z}\right] + z^2 \right\}
\\
= & \,\delta(1-z)  \, \frac{\alpha_s}{2\pi}\,C_F \, 
\left(\frac{4\pi\mu^2}{Q^2}\right)^{-\epsilon} 
\frac{\Gamma(1+\epsilon)}{\Gamma(1+2\epsilon)}\,
\left\{-\frac{8}{3}\frac{1}{\epsilon} + \frac{5}{9} \right\}.
\end{split}
\label{eq:GamV2qqMC}
\end{equation} 
Thus the full virtual correction reads
\begin{equation}
\hat{\Gamma}^{\mc}_{q\bar{q},V}(z,\epsilon)  = 
\delta(1-z)  \, \frac{\alpha_s}{2\pi}\,C_F \, \left(\frac{4\pi\mu^2}{Q^2}\right)^{-\epsilon} 
\frac{\Gamma(1+\epsilon)}{\Gamma(1+2\epsilon)}\,
\left\{-\frac{2}{\epsilon^2} + \frac{3}{\epsilon} - \frac{17}{2} - \frac{2\pi^2}{3} \right\}.
\label{eq:GamVqqMCfull}
\end{equation} 
After combining it with the real correction of Eq.~(\ref{eq:GamRqqMC}) we obtain a complete 
MC radiation function:
\begin{equation} 
\begin{split}
&\hat{\Gamma}^{\mc}_{q\bar{q}}(z,\epsilon) =
\frac{\alpha_s}{2\pi}C_F \left(\frac{4\pi\mu^2}{Q^2}\right)^{-\epsilon} 
\frac{\Gamma(1+\epsilon)}{\Gamma(1+2\epsilon)}
\Bigg\{
 \frac{2}{\epsilon} \left[\frac{1+z^2}{(1-z)_+} + \frac{3}{2}\delta(1-z) \right]
 \\ &~~~~
 - \delta(1-z) \left(\frac{2\pi^2}{3} + \frac{17}{2}\right)
 + 4(1+z^2) \left[\frac{\ln(1-z)}{1-z}\right]_+ - 2 \frac{1+z^2}{1-z}\ln z + 2(1-z)
 \Bigg\}.
\end{split}
\label{eq:GamqqMC}
\end{equation} 

The corresponding  NLO correction reads \cite{Altarelli:1979ub}
\begin{equation}
\begin{split}
\hat{\rho}^{\nlo}_{q\bar{q}}(z,\epsilon) 
= &\frac{\alpha_s}{2\pi}\,C_F \, 
 \left(\frac{4\pi\mu^2}{Q^2}\right)^{-\epsilon} 
 \frac{\Gamma(1+\epsilon)}{\Gamma(1+2\epsilon)}\,
\Bigg\{
 \frac{2}{\epsilon} \left[\frac{1+z^2}{(1-z)_+} + \frac{3}{2}\delta(1-z) \right]
\\ & 
 - \delta(1-z) \left( \frac{2\pi^2}{3} - 8 \right)
 + 4(1+z^2) \left[\frac{\ln(1-z)}{1-z}\right]_+ - 2 \frac{1+z^2}{1-z}\ln z 
 \Bigg\}.
\end{split}
\label{eq:rhoqqNLO}
\end{equation}
Then, for the coefficient function in the MC factorisation scheme we obtain
\begin{equation}
 C_{q\bar{q}}^{\rm MC}(z) 
 = \hat{\rho}^{\nlo}_{q\bar{q}}(z,\epsilon) - \hat{\Gamma}^{\mc}_{q\bar{q}}(z,\epsilon)
 = \frac{\alpha_s}{2\pi}\, C_F \left\{ \delta(1-z) 
   \left(\frac{4\pi^2}{3} + \frac{1}{2}\right) - 2(1-z) \right\}.
\label{eq:C2qqMC}
\end{equation}
The above expression differs from the one given in Ref.~\cite{Jadach:2015mza},
\begin{equation}
C_{2q}^{\rm MC}(z) = \frac{\alpha_s}{2\pi}\, C_F 
\left\{\delta(1-z) \left(\frac{4\pi^2}{3} - \frac{5}{2}\right) - 2(1-z) \right\} ,
\label{eq:C2qMC-old}
\end{equation}
by a constant term:
\begin{equation}
C_{q\bar{q}}^{\rm MC}(z) -  C_{2q}^{\rm MC}(z) = \frac{3 C_F\alpha_s}{2\pi}\,\delta(1-z)\,.
\label{eq:C2qMC}
\end{equation}

For completeness, let us also write the corresponding 
coefficient function in the $\msbar$ 
factorisation scheme:
\begin{equation}
C_{q\bar{q}}^{\overline{\rm MS}}(z) 
= \frac{\alpha_s}{2\pi}\,C_F \,
 \left\{ \delta(1-z) \left(\frac{4\pi^2}{3} 
- \frac{7}{2}\right) + \left[2\frac{1 + z^2}{1-z} \ln \frac{(1-z)^2}{z}\right]_+
\right\}
\label{eq:C2qqMSbar}
\end{equation}
and the $qq$ transformation matrix element to the quark PDF in the MC scheme:
\begin{equation}
K^{\rm MC}_{qq}(z) = 
\frac{1}{2} \left[C_{q\bar{q}}^{\overline{\rm MS}}(z) 
- C_{q\bar{q}}^{\rm MC}(z) \right]
= \frac{\alpha_s}{2\pi}\,C_F \,
 \left\{ \left[ \frac{1 + z^2}{1-z} \ln \frac{(1-z)^2}{z} 
 + 1 - z \right]_+  - \frac{3}{2}\delta(1 - z) \right\}.
\label{eq:DeltaC2qq1}
\end{equation}
This can also be expressed in a form similar 
to the corresponding formula for the $gg$ channel,
cf.\ Eq.~(\ref{eq:DeltaC2gg}):
\begin{equation}
\begin{split}
K^{\rm MC}_{qq}(z)
= \frac{\alpha_s}{2\pi}\,C_F \,
 \Bigg\{  &
 4\left[\frac{\ln(1-z)}{1-z}\right]_+ 
 - (1+z) \ln \frac{(1-z)^2}{z} - 2 \frac{\ln z}{1-z} + 1 - z   
 \\ &
 - \delta(1 - z)\left(\frac{\pi^2}{3} + \frac{17}{4}\right)
 \Bigg\}.
\end{split}
\label{eq:DeltaC2qq2}
\end{equation}
This is the $q\rightarrow q$ PDF transition-matrix element 
from the  $\overline{\rm MS}$ to MC scheme. 
Similarly as in the previous cases, 
it can also be obtained from the respective counter terms:
\begin{equation}
K^{\rm MC}_{qq}(z) = 
\left[ \hat{\Lambda}_{qq}^{\rm MC}(z,\epsilon) 
     - \hat{\Lambda}_{qq}^{\overline{\rm MS}}(z,\epsilon) 
\right]_{\epsilon = 0},
\label{eq:DeltaC2qqCC}
\end{equation}
where the universal MC counter term corresponding to the $q\to q$ transition can be related
to the MC radiation function 
$\hat{\Gamma}_{q\bar{q}}^{\rm MC}(z,\epsilon)$ of Eq.~(\ref{eq:GamqqMC}):
\begin{equation}
\hat{\Lambda}_{qq}^{\rm MC}(z,\epsilon)  = \frac{1}{2} \hat{\Gamma}_{q\bar{q}}^{\rm MC}(z,\epsilon),
\label{eq:CCqqMC}
\end{equation}
while
\begin{equation}
\hat{\Lambda}_{qq}^{\overline{\rm MS}}(z,\epsilon) =
\frac{\alpha_s}{2\pi}C_F \left(\frac{4\pi\mu^2}{Q^2}\right)^{-\epsilon} 
\frac{\Gamma(1+\epsilon)}{\Gamma(1+2\epsilon)}\,
 \frac{1}{\epsilon} \left[\frac{1+z^2}{(1-z)_+} + \frac{3}{2}\delta(1-z) \right]
\label{eq:GamqqMSbar}
\end{equation}
is the corresponding $\msbar$ counter term. 

\section{PDFs in MC scheme}
\label{subsec:PDFMC}

In Ref.~\cite{Jadach:2015mza}, were the  \krknlo{} method was applied to the
Drell-Yan process, it was 
sufficient to transform the $\msbar$ PDF of quarks and antiquarks.
The difference between the $\msbar$ and MC PDFs for the gluon was an NNLO effect,
hence beyond the claimed accuracy.

Here, for the Higgs production process, the gluon PDF also has to be transformed to
the MC scheme.
Having calculated all the necessary ingredients in the previous section,
we define this transformation as follows:
\begin{equation}
g_{\rm MC}(x,Q^2) 
= g_{\overline{\rm MS}}(x,Q^2)\; + \int_x^1 \frac{dz}{z}\, 
  g_{\overline{\rm MS}} \left(\frac{x}{z},Q^2\right) K^{\rm MC}_{gg}(z) \; 
+ \sum_q  \int_x^1 \frac{dz}{z}\, 
  q_{\overline{\rm MS}} \left(\frac{x}{z},Q^2\right) K^{\rm MC}_{gq}(z),
\label{eq:gluonPDF}
\end{equation}
where $K^{\rm MC}_{gg}(z)$ is given in Eq.~(\ref{eq:DeltaC2gg}) 
and $K^{\rm MC}_{gq}(z)$ in Eq.~(\ref{eq:DeltaC2gq}).
However, virtual parts of the transformation matrix in the quark sector 
now has also changed due to the necessary use of the momentum sum rules.
Hence, the entire transformation rule now takes the form
\begin{equation}
  \begin{bmatrix}  
     q(x,Q^2) \\ \bar{q}(x,Q^2) \\ g(x,Q^2) 
  \end{bmatrix}_{\mc}
=
 \begin{bmatrix}  
     q \\ \bar{q} \\ g 
 \end{bmatrix}_{\msbar}
+
\int\! dz dy
 \begin{bmatrix}
    K^{\mc}_{qq}(z)     & 0                       &  K^{\mc}_{qg}(z)  \\ 
    0                 & K^{\mc}_{\bar{q}\bar{q}}(z)  &  K^{\mc}_{\bar{q}g}(z) \\ 
    K^{\mc}_{gq}(z)     & K^{\mc}_{g\bar{q}}(z)       &  K^{\mc}_{gg}(z)
 \end{bmatrix}
 \begin{bmatrix}  
     q(y,Q^2) \\ \bar{q}(y,Q^2) \\ g(y,Q^2) 
 \end{bmatrix}_{\msbar}
  \!\!\!\!\!\!   \delta(x-yz),
\label{eq:transPDF}
\end{equation}
where
\begin{equation}
\begin{split}
 K^{\mc}_{gq}(z)
&= \frac{\alpha_s}{2\pi}\, C_F \left\{ \frac{1 + (1-z)^2}{z}\ln\frac{(1-z)^2}{z} + z\right\},
\\
 K^{\mc}_{gg}(z) 
&=\frac{\alpha_s}{2\pi}\, C_A 
 \Bigg\{
  4\left[\frac{\ln(1-z)}{1-z} \right]_+ 
  + 2\left[\frac{1}{z} - 2 + z(1-z) \right]\ln\frac{(1-z)^2}{z}  
\\&~~~~~~~~~~~~~~~~~~~~~~~~~~
  - 2 \frac{\ln z}{1-z} 
  - \delta(1-z) \left(  \frac{\pi^2}{3} + \frac{341}{72} 
                      - \frac{59}{36}\frac{T_f}{C_A}\right)
 \Bigg\},
\\
K^{\rm MC}_{qq}(z)
&= \frac{\alpha_s}{2\pi}\,C_F \,
 \Bigg\{
   4\left[\frac{\ln(1-z)}{1-z}\right]_+ 
   - (1+z) \ln \frac{(1-z)^2}{z} - 2 \frac{\ln z}{1-z} + 1 - z
\\&~~~~~~~~~~~~~~~~~~~~~~~~~~
   - \delta(1 - z)\left(\frac{\pi^2}{3} + \frac{17}{4}\right)
 \Bigg\},
\\
K^{\rm MC}_{qg}(z)
&= \frac{\alpha_s}{2\pi}\,T_R \,
 \left\{ \left[z^2 + (1-z)^2\right]\ln \frac{(1-z)^2}{z}  + 2z(1-z) \right\},
 \\
 K^{\mc}_{g\bar{q}}(z)
&= K^{\mc}_{gq}(z),
\qquad
K^{\mc}_{\bar{q}g}(z)
= K^{\mc}_{qg}(z).
 \end{split}
\label{eq:Kmatrix}
\end{equation}
The above formulae can be used for 
numerical computation of the MC-scheme quark and gluon PDFs
from the available parametrisation of the $\overline{\rm MS}$ PDFs.
Alternatively, PDFs in the MC scheme can be fitted directly to DIS and other
data, provided the NLO coefficient functions in the MC scheme are known.
For DIS they are listed in Appendix~\ref{appendixA}.

We assume that PDFs in the MC scheme satisfy the same momentum sum rule 
as PDFs in the $\msbar$ scheme:
\begin{equation}
\int_0^1 dx \; x 
  \Big[ g_{\rm MC}(x,Q^2) + \sum_q  q_{\rm MC}(x,Q^2) 
  \Big] 
= \int_0^1 dx \; x 
  \Big[ g_{\overline{\rm MS}}(x,Q^2) + \sum_q q_{\overline{\rm MS}}(x,Q^2) 
  \Big].
\label{eq:PDFs-MSR}
\end{equation}
Inserting in the above formula the expressions for $g_{\rm MC}$ and $q_{\rm MC}$ 
from Eqs.~(\ref{eq:transPDF}), 
we obtain the following momentum sum rules
for the factorisation-scheme transformation matrix elements:
\begin{equation}
\begin{split}
&  \int_0^1 dz \; z \left[  K^{\rm MC}_{qq}(z) +   K^{\rm MC}_{gq}(z) \right] =  0\,,
\\
& \int_0^1  dz \; z \left[  K^{\rm MC}_{gg}(z) +   2n_f K^{\rm MC}_{qg}(z) \right]  =  0\,.
\label{eq:Cg-MSR}
\end{split}
\end{equation}
The above, of course, results from the momentum sum rules imposed on
the MC and $\msbar$ soft-collinear counter terms, 
however it constitutes a useful cross-check of the consistency of the MC scheme.


Looking at the elements of the transition-matrix $K$ in Eq.~(\ref{eq:Kmatrix}) one can see that
the terms $\sim \ln(1-z)$ and $\sim \ln z$ are absorbed in the MC-scheme PDFs. 
As a result the NLO coefficient functions for the DY process and the Higgs-boson production
are much simpler than the corresponding ones in the $\msbar$ scheme, 
cf. Eqs.~(\ref{eq:C2qgMC}) and (\ref{eq:C2qgMSbar}), (\ref{eq:C2qqMC}) and (\ref{eq:C2qqMSbar}),
(\ref{eq:C2ggMC}) and  (\ref{eq:C2ggMSbar}), (\ref{eq:C2gqMC}) and (\ref{eq:C2gqMSbar}).
One can thus expect that higher-order QCD corrections, beyond NLO, will be smaller 
in the MC factorisation scheme than in the $\msbar$ scheme.  
In particular, the MC-scheme coefficient functions are free of the so-called leading threshold corrections,
$\sim \ln(1-z)/(1-z)$, which are absorbed (and resummed) in the MC PDFs. 

Let us summarise on the motivation of introducing the new, MC PDFs and their main
features, in form of a list of questions and answers:
\begin{itemize}
\item
What is the purpose of MC factorisation scheme?
It is defined such that the $\Sigma(z)\delta(k_T)$ terms due to emission
from initial partons disappear completely from the real NLO corrections
in the exclusive/unintegrated form,
even before PSMC gets involved.
\item
Why is the above vital in the \krknlo{} scheme?
Without eliminating such terms
it is not possible to include the NLO corrections
using simple multiplicative MC weights on top of distributions
generated by PSMC.
\item
How to determine elements of the transition matrix $K^{\mc}_{ab}$?
They can be deduced from the difference of soft-collinear counter terms
of the MC and $\msbar$ scheme or from inspection
of the NLO corrections in a few simple processes
with initial quarks and gluons in the LO hard process.
We have done it both ways.
\item
Will the same PDFs in the MC scheme eliminate $\sim\delta(k_T)$ terms for all processes?
This is a question about the {\em universality} of the MC factorisation scheme.
For all processes similar to the DY or Higgs-production process, with produced colour-neutral
final-state objects, the answer is positive.
\end{itemize}

\begin{figure}[!t]
  \centering
  \includegraphics[width=0.47\textwidth]{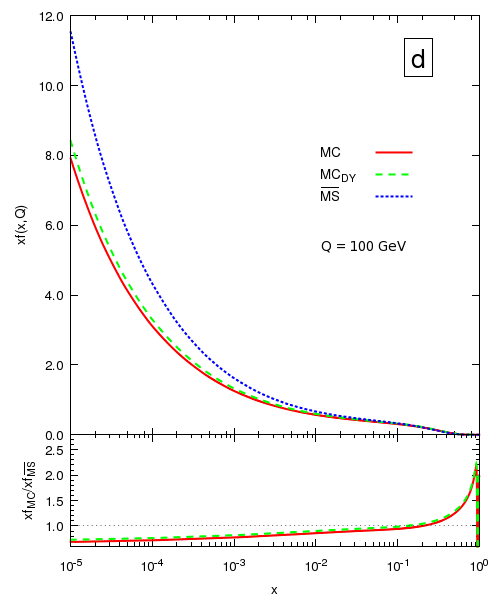}
  \hfill
  \includegraphics[width=0.47\textwidth]{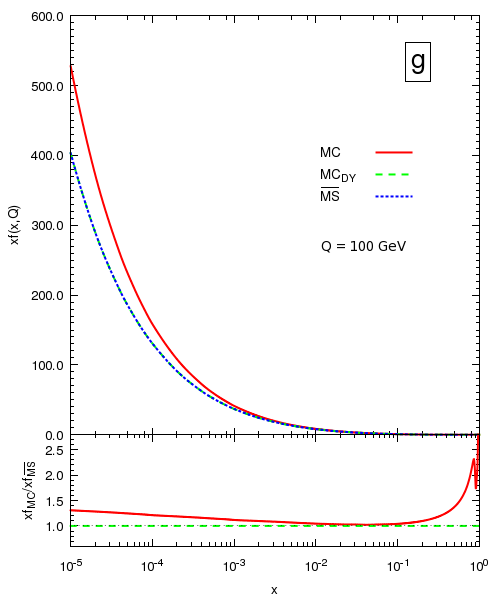}
  \includegraphics[width=0.47\textwidth]{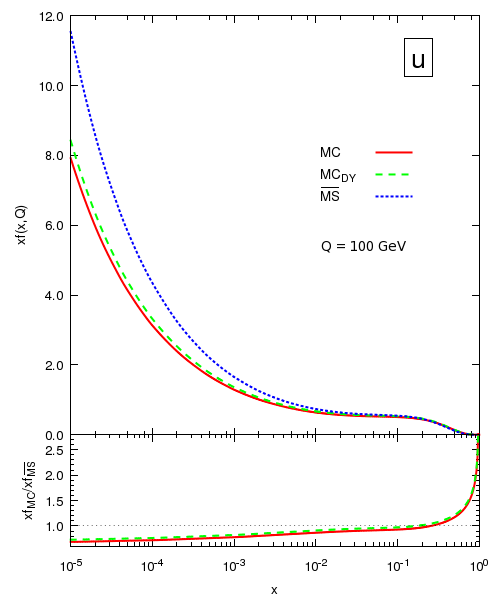}
  \hfill
  \includegraphics[width=0.47\textwidth]{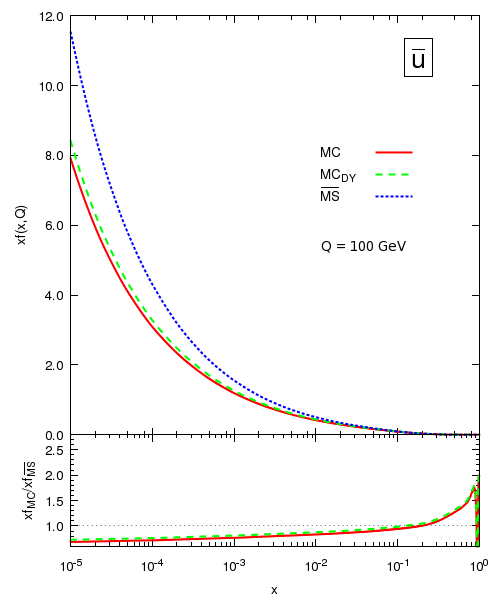}
  \caption{\sf
  Comparison of PDFs in the MC and $\msbar$ factorisation schemes. 
  PDFs denoted with ${\sf MC}_{\sf DY}$ are the ones used for the Drell--Yan process in Ref.~\cite{Jadach:2015mza}.
  }
  \label{fig:mcPDFs}
\end{figure}

In Fig.~\ref{fig:mcPDFs}, we present
examples of numerical results for the PDFs of quarks and gluon in the MC
scheme obtained from PDFs in $\msbar$ scheme
using transformation of Eqs.~(\ref{eq:transPDF}) and~(\ref{eq:Kmatrix}).
The upper panels show the absolute values of the $\msbar$ and MC parton
distributions taken at the scale $Q=100\, \GeV$, whereas the ratios of the two
are displayed in the lower panels.

Two types of MC PDFs are plotted: the complete version (red solid), where both
quarks and gluons are transformed, and the ``DY'' version (green dashed), where
the gluon is unchanged with respect to $\msbar$. As discussed earlier, these
types of MC PDFs is sufficient for the Drell--Yan process and it was used in our
previous work~\cite{Jadach:2015mza}. Hence, we show them here for comparison.

One can see that the differences between the MC and $\msbar$ PDFs are noticeable. In
particular, the MC quarks are up to $20\%$ smaller at low and moderate $x$, while
they get above the $\msbar$ distributions at large $x$. For DY and Higgs production,
the latter has consequences only at large rapidities of the bosons.
At the same time, we notice that the gluon is larger in the MC scheme at low and
moderate $x$. Hence, the changes in quarks and the gluon have a chance to
compensate each other and, indeed, as we checked explicitly, the momentum sum rules
(\ref{eq:PDFs-MSR}) are numerically satisfied for our MC PDFs.
 
Other quark flavours, when transformed to the MC scheme, exhibit similar changes
to those shown in Fig.~\ref{fig:mcPDFs} for the $u$ and $d$ quarks.

Finally, let us comment briefly on the process-independence (universality)
of the MC factorisation scheme and the \krknlo{} method.
If we treat Eq.~(\ref{eq:transPDF}) as a definition of PDFs in the MC scheme,
then their universality is just inherited from the $\msbar$ scheme.
The universality of the \krknlo{} method is more involved and it would imply
that by means of adoption of these PDFs and a careful choice of 
the exclusive/unintegrated MC distributions for the initial-state splittings,
we are able to eliminate from the NLO real corrections all terms
proportional to $\delta(\beta) f(z)$ or $\delta(\alpha) f(z)$,
which means that we can impose the NLO real corrections with the multiplicative
MC weights in $d=4$ dimensions on top of the PSMC distributions.
We are able to state that the above is true for all annihilation process into
colour-neutral objects.
This can be deduced from analysing the CS counter terms
(which are compatible with the modern PSMCs),
where both the emitter and spectator are in the initial state.
They are universal within the class of the above annihilation processes
and therefore the \krknlo{} method features the same property.
The answer to the question whether extending this argument to other processes,
with one or more coloured partons in the final state at the LO level,
is not trivial and the relevant study is reserved
to next dedicated publication%
\footnote{The analysis in Ref.~\cite{Jadach:2011cr} 
 for the DIS process, albeit limited to the gluonstrahlung
 NLO subprocess, gives hope for a possible positive answer.}.

\section{NLO cross sections for Higgs production in \krknlo{} method}

MC weights of the \krknlo{} method for the Higgs-boson production in gluon--gluon fusion are very simple,
even simpler than those for Drell-Yan process, where they depend
on the angles of the $Z/\gamma^*$ decay products.
For the $g + g \rightarrow H + g$ subprocess we have
\begin{equation}
W_R^{gg}(\alpha,\beta) 
= \frac{|{\cal M}_{gg\rightarrow Hg}^{\nlo}|^2}{|{\cal M}_{gg\rightarrow Hg}^{\mc}|^2} = 
 \frac{1 + z^4 + \alpha^4 + \beta^4}{2\left[ z^2 + (1-z)^2 + z^2(1-z)^2\right]} =
 \frac{1+z^4 + \alpha^4 + \beta^4}{1 + z^4 + (1 - z)^4} \quad \leq 1,
\label{eq:wt-gg}
\end{equation}
whereas for  the $g + q \rightarrow H + q$ channel, the real weight reads
\begin{equation}
W_R^{gq}(\alpha,\beta) 
= \frac{|{\cal M}_{gq\rightarrow Hq}^{\nlo}|^2}{|{\cal M}_{gq\rightarrow Hq}^{\mc}|^2} = 
\frac{1 + \beta^2}{1 + (1-z)^2}  \quad \leq 1\,.
\label{eq:wt-gq}
\end{equation}
For the process with exchanged initial-state partons we have
$W_R^{qg}(\alpha,\beta) = W_R^{gq}(\beta,\alpha)$. 

Virtual+soft-real corrections 
can be read off from the formulae of the
coefficient functions given in Section~\ref{sec:MCscheme}.
They are just constant terms multiplied
by the $\delta(1-z)$ function. 
In the \krknlo\ method they 
should be included multiplicatively in 
a parton shower generator for the corresponding process,
i.e.\ the Born-level cross section should be multiplied by the weight 
\begin{equation}
W_{VS} = 1 + \Delta_{VS} ,  
 \label{eq:VirWgt}
\end{equation}
where $\Delta_{VS}$ is the virtual+soft-real correction.
For the Higgs-boson production, it can be read off from Eqs.~(\ref{eq:C2ggMC}) 
and (\ref{eq:C2gqMC}) to get
\begin{equation}
 \Delta^{gg}_{VS} = \frac{\alpha_s}{2\pi}\, C_A
 \left(\frac{4\pi^2}{3} + \frac{473}{36} - \frac{59}{18} \frac{T_f}{C_A}\right),
 \qquad 
 \Delta^{gq}_{VS} = 0.
 \label{eq:VirggHMC}
\end{equation}

For the numerical evaluation of the cross sections
at the LHC 
for the proton--proton collision energy of $\sqrt{s}=8$~TeV,
we choose the following set of the Standard Model (SM) input parameters:
\begin{eqnarray}\label{eq:pars}
M_H = 126 \; {\rm GeV}, & \quad & \Gamma_Z =  2.4952  \; {\rm GeV},
\nonumber  \\
M_W = 80.4030 \; {\rm GeV}, & \quad & \Gamma_W = 2.1240 \; {\rm GeV},
\\
M_Z = 91.1876 \; {\rm GeV}, & \quad & \alpha_s(M_Z^2)=0.13938690
\\
G_{\mu} = 1.16637\times 10^{-5} \; {\rm GeV}^{-2}, 
& \quad & m_t = 173.2  \; {\rm GeV},
\nonumber  
\end{eqnarray}
and the $G_{\mu}$-scheme~\cite{LHC-YR} for the electroweak sector. 
To compute the hadronic cross section we also use the {\tt MSTW2008} LO set of parton
distribution functions~\cite{Martin:2009iq}, and take the renormalisation and
the factorisation scales to be $\mu_R^2=\mu_{F}^2=M_H^2$, where $M_H$ is the Higgs-boson mass.
We also set the Higgs boson to be stable for simplicity. 
\begin{table}[!h]
\centering
  \begin{tabular}{|l|c|}
 \hline 
         \multicolumn{2}{|c|}{$\sigma^{tot}_\text{H}$ [pb]}                   
   \\ \hline \hline
  \mcatnlo{}        &   $18.72  \pm  0.04$     
   \\ \hline 
  \krknlo{}         &  $19.38 \pm 0.04$    
  \\ \hline 
  \end{tabular}
  \caption{\sf
   Values of the total cross section with statistical errors for the Higgs-boson production in gluon--gluon fusion 
   at NLO from the \krknlo{} method compared to the results of \mcatnlo{}.
  }
  \label{tab:krknlo-H-xsectio}
\end{table}
In Table~\ref{tab:krknlo-H-xsectio} we show
the results for the total cross sections for the Higgs-boson production in gluon--gluon fusion
obtained with \krknlo{} and 
\mcatnlo{}. The two methods are matched to the dipole 
parton shower implemented in 
\herwig{7}~\cite{Platzer:2011bc, Bellm:2015jjp}.

We see that the two methods give slightly different ($\sim 3.5\%$) total cross sections, which
come from formally higher-order terms, i.e. beyond the NLO approximation. The relevant distributions and detailed 
comparisons with \mcatnlo{}, \powheg{} and NNLO calculations will be presented in another 
publication~\cite{IFJPAN-IV-2016-10}.

\section{Summary and outlook}
In this work, we have presented all the ingredients of the \krknlo{} method
needed 
for its implementation for the Higgs-boson production process in gluon--gluon fusion.
In particular, the complete definitions of PDFs in the MC scheme, together with their
numerical distributions, have been provided. Hence, PDFs in the MC FS can be fitted 
directly to experimental DIS and DY data.
We have also presented the first result for the total cross section for the Higgs
production.
More distributions, comparisons with \mcatnlo{}, \powheg{} and NNLO calculations
will be presented in another publication~\cite{IFJPAN-IV-2016-10}.
A dedicated study of the process-independence (universality) of the \krknlo{} method
and the MC factorisation scheme is also reserved for the future work.

\section*{\large Acknowledgements}
We are grateful to Simon Platzer and Graeme Nail for the useful discussions and their help 
with the dipole parton shower implemented in \herwig{7}.
We are indebted to the Cloud Computing for Science and Economy 
project (CC1) at IFJ PAN (POIG 02.03.03-00-033/09-04) in Krak\'ow whose resources 
were used to carry out some of the numerical computations for this project. 
We also thank Mariusz Witek and Mi{\l}osz Zdyba{\l} for their help with CC1. 
This work was funded in part by 
the MCnetITN FP7 Marie Curie Initial Training Network PITN-GA-2012-315877.

\appendix
\section{Coefficient functions for DIS process in MC scheme}
\label{appendixA}
The NLO coefficient functions $C_2$  
for deep-inelastic electron--proton scattering (DIS) in the
$\overline{\rm MS}$ factorisation scheme read
\begin{eqnarray}
C_{2,qq}^{\overline{\rm MS}}(z) &=& \frac{\alpha_s}{2\pi}\,C_F \,
\left[ \frac{1+z^2}{1-z}\ln\frac{1-z}{z} - \frac{3}{2} \frac{1}{1-z} + 2z + 3\right]_+,
\label{eq:C2qqDIS-MSb}\\
C_{2,qg}^{\overline{\rm MS}}(z) &=& \frac{\alpha_s}{2\pi}\,T_R \,
\left\{ \left[ z^2 + (1 - z)^2 \right] \ln\frac{1-z}{z}  + 8z(1-z) - 1 \right\}. 
\label{eq:C2qgDIS-MSb}
\end{eqnarray}

The corresponding coefficient functions in the 
MC factorisation scheme can be obtained from the
above formulae with the help of the transformation matrix elements 
$ K^{\rm MC}_{ij}$ in the
following way:
\begin{eqnarray}
C_{2,qq}^{\rm MC}(z) &=& C_{2,qq}^{\overline{\rm MS}}(z) -  K^{\rm MC}_{qq}(z) 
\nonumber\\
&=&
\frac{\alpha_s}{2\pi}\,C_F \,
\left\{\left[ -\frac{1+z^2}{1-z}\ln(1-z) 
- \frac{3}{2} \frac{1}{1-z} + 3z + 2\right]_+ + \frac{3}{2}\,\delta(1-z)\right\},
\label{eq:C2qqDIS-MC}\\
C_{2,qg}^{\rm MC}(z) &=& C_{2,qg}^{\overline{\rm MS}}(z) -  K^{\rm MC}_{qg}(z) 
\nonumber\\
&=& \frac{\alpha_s}{2\pi}\,T_R \,
\left\{ -\left[ z^2 + (1 - z)^2 \right] \ln(1-z)  + 6z(1-z)  - 1 \right\}. 
\label{eq:C2qgDIS-MC}
\end{eqnarray}
These coefficient functions can be used in fitting the MC PDFs to experimental DIS data.



\providecommand{\href}[2]{#2}

\end{document}